\documentclass[aps,amsfonts,nofootinbib]{revtex4-1}
\usepackage{epsfig}
\usepackage{feynmp}
\usepackage{graphicx}
\usepackage{slashed}
\usepackage{amsmath}
\usepackage{amsbsy}
\usepackage{bm}
\usepackage{color}
\usepackage{epsfig}
\usepackage{ulem}
\newcommand{\ee}{\end{equation}}
\newcommand{\bb}{\begin{equation}}
\newcommand{\eqb}{\begin{eqnarray}}
\newcommand{\eqf}{\end{eqnarray}}

\newcommand{\1}{{\'{\i}}}

\def\1{\'{\i}}

\def\1{\'{\i}}
\begin{document}
\title{Higgs Model Coupled to Dark Photons}
\author{D. Carcamo}
\email{dante.carcamo@usach.cl}
\address{Departamento de F\1sica, Universidad de Santiago de Chile, Casilla 307, Santiago, Chile}
\author{J. Gamboa}
\address{Departamento de F\1sica, Universidad de Santiago de Chile, Casilla 307, Santiago, Chile}
\author{M. Pino}
\email{miguel.pino.r@usach.cl}
\address{Departamento de F\1sica, Universidad de Santiago de Chile, Casilla 307, Santiago, Chile}

\begin{abstract}
A dark sector containing a Higgs field and a dark photon coupled to visible photons via a small kinetic mixing is considered. The mass of the dark gauge boson becomes rescaled 
by the parameter involved in the kinetic mixing.  We also calculate the total annihilation cross section for scalar Higgs interacting with an external field and show that exhibits Sommerfeld enhancement. 
The model can be extended to a  $SU(3)\times SU(2)\times U(1)$ dark sector. The mass of the dark $Z$ boson increases due to the kinetic mixing.   
\end{abstract}
\pacs{PACS numbers:}
\date{\today}
\maketitle

\section*{Introduction}

Since the seventies it has been observed that the gamma ray spectrum coming from the galactic center possesses a 511 keV emission line consistent with a copious annihilation of electron-positron pairs \cite{jo,le,sha,ki,atti,ve,kn,wei,tee}. Nonetheless, the calculated rate of $\sim 10^{42}$ annihilations per second exceeds in several order of magnitude the creation rate of electron-positron pairs from the interaction of cosmic rays with the interstellar medium. This and other anomalies in the cosmic ray observations (see \cite{arkani} for a summary) could be naturally explained by assuming that the overabundance of matter-antimatter pairs is due to an (Sommerfeld) enhancement of the total cross section of dark matter annihilation. This opens a possibility for new physics in order to address the question of how ordinary and dark matter interact to produce this enhancement.

Several routes have been proposed to explain this puzzle (see {\it e.g.} \cite{arkani}, \cite{pospe} or \cite{review0} for recent references), although  the most popular approaches to dark matter are WIMPS's and new interactions beyond the standard model \cite{beyond}.
 
In this paper we propose a simple model containing a hidden sector composed of a dark Higgs field and a dark photon coupled with visible photons through a small kinetic mixing \cite{holdom}. In the first section, we show that this interaction is enough to produce an enhancement of electromagnetic processes in the dark sector. In the second part, we analyze the consequences of kinetic mixing in an extended dark model $SU (3) \times SU (2)\times U_Y (1) \times U_{Y'}(1) $, where $ U_{Y'} (1) $ is the visible hypercharge. The mass of the dark vector boson $Z$ turns out to be bigger than the usual $ Z $.

\section{A Toy Model}

This first part is devoted to the problem of determining the total annihilation cross section for a Higgs model coupled to dark photons, explicitly showing the enhancement. 
For concreteness, let us start considering the following model   
\bb 
{\cal L} = {\cal L}_{Dark} + {\cal L}_{I}, \label{model1}
\ee 
where ${\cal L}_{Dark}$ is a scalar electrodynamics-like Lagrangian for the hidden sector
\bb 
{\cal L}_{Dark} = |D(A) \phi|^2 - m^2 |\phi|^2 -\frac{\lambda}{2} (|\phi|^2)^2-\frac{1}{4e^2_A} F_{\mu \nu} (A) F^{\mu \nu} (A), \label{mod2} 
\ee 
where $A_\mu$ and $\phi$ are dark fields and $D_\mu (A) = \partial_\mu -i A_\mu$. Now suppose that at some energy scale an interaction with the visible photon appears. The Lagrangian ${\cal L}_{I}$ is then given by  
\bb 
{\cal L}_{I} =  - \frac{1}{4e^2_B} F_{\mu \nu} (B) F^{\mu \nu} (B) + \frac{\gamma}{2} F_{\mu \nu} (A) F^{\mu \nu} (B),  \label{kin1}
\ee  
where  $B_\mu$ is the visible photon, $F_{\mu \nu} (A) F^{\mu \nu} (B)$ is the kinetic mixing and $\gamma$ is a small dimensionless coefficient \cite{ring}. 

The quantum theory is described by the integral 
\bb 
{\cal Z} = \int {\cal D} \phi {\cal D} \phi^* \int \left[ {\cal D} A_\mu {\cal D} B_\mu \right] \, e^{ i S[\phi, \phi^*,  A, B]} \label{int}
\ee 
where $ \left[ {\cal D} A_\mu {\cal D} B_\mu \right]$ is a compact notation that includes  the gauge fixing and the Faddev-Popov determinant. 

In order to obtain a mixture of visible and dark coupling constant, it is convenient to perform the change of variables
\eqb 
A'_\mu &=& A_\mu \nonumber, 
\\
B'_\mu &=& B_\mu - \gamma e^2_B A_\mu, \nonumber
\eqf 
which diagonalizes ${\cal L}_{I}-\frac{1}{4e^2_A} F_{\mu \nu} (A) F^{\mu \nu} (A)$ as follows 
\bb 
 -\frac{1}{4 {\tilde e}^2} F^2 (A')-\frac{1}{4 e^2_B} F^2 (B'),  \label{dia}
\ee
but paying the price of redefining the electric charge $e_A$ by the  effective charge ${\tilde e}$, {\it i.e.}
\bb 
{\tilde e} = \frac{e_A}{\sqrt{1-(\gamma e_B e_A)^2}}. 
\ee

In other words, (\ref{model1}) must be written as \footnote{ Note that, for simplicity, we drop the quotes and write $A,B$ instead of $A',B'$.}
\bb 
{\cal L} = |D(A) \phi|^2 - m^2 |\phi|^2 - \frac{\lambda}{2} (|\phi|^2)^2 -\frac{1}{4 {\tilde e}^2} F^2 (A) - \frac{1}{4 e^2_B} F^2 (B). \label{int}
\ee

The path integral becomes
\eqb
{\cal Z} &=&  \int {\cal D} \phi {\cal D} \phi^* \int \left[ {\cal D} A_\mu {\cal D} B_\mu \right] \, e^{i \int d^4x\left( |D(A) \phi|^2 - m^2 |\phi|^2 - \frac{\lambda}{2} (|\phi|^2)^2 -\frac{1}{4 {\tilde e}^2} F^2 (A) - \frac{1}{4 e^2_B} F^2 (B) \right)} \nonumber 
\\
&=& {\cal N} \int {\cal D} \phi {\cal D} \phi^* \int {\cal D} A_\mu \,  e^{i \int d^4x\left( |D(A) \phi|^2 - m^2 |\phi|^2 - \frac{\lambda}{2} (|\phi|^2)^2 -\frac{1}{4 {\tilde e}^2} F^2 (A) \right)} \nonumber 
\\ 
&=&  {\cal N} \int {\cal D} \phi {\cal D} \phi^* \int {\cal D} A_\mu \,  e^{i \int d^4x{\cal L}}, \label{int2}
\eqf
where the $B_\mu$ field have been absorbed in the normalization constant ${\cal N }$, leaving the effective action as 
\bb
{\cal L}_{eff}=  |D(A) \phi|^2 - m^2 |\phi|^2 - \frac{\lambda}{2} (|\phi|^2)^2 -\frac{1}{4 {\tilde e}^2} F^2 (A), \label{la1}
\ee 
which is a Higgs model with a nontrivial rescaling of the electric charge. Furthermore, we can hide the charge $\tilde e$ by redefining the potential $A_\mu$  as   
\bb
A_\mu  \to  {{\tilde e}}{ A}_\mu, \label{refe}
\ee 
so that the effective Lagrangian (\ref{la1}) becomes
\bb 
{\cal L}_{eff} = |D(A) \phi|^2 - m^2 |\phi|^2 - \frac{\lambda}{2} (|\phi|^2)^2 -\frac{1}{4} F^2 (A), \label{kin44}
\ee 
where  the redefined covariant derivative is   
\[ 
D_\mu [A] = \partial_\mu -i{\tilde e}A_\mu.  
\]

Therefore, a first conclusion is that the kinetic mixing with the visible photon is equivalent to shield the electric charge $e_A$. The effective model is just an abelian Higgs model with the coupling constant ${\tilde e}$ instead of $e$. This conclusion is valid not only for the Higgs field. Any other particle in the dark sector coupled to the dark photon would have its charge redefined in the same way.

Now we can perform spontaneous symmetry breaking. Using the Kibble parametrization, the scalar field can be written as 
\bb 
\phi = \frac{1}{\sqrt{2}} \left( v + h \right) \, e^{\frac{i}{v} \xi}, \label{higgs}
\ee 
where $v$ is the minimum of the potential, {\it i.e.} 
\bb
v= \sqrt{-\frac{2m^2}{\lambda}}, \label{vev}
\ee
and  $h$ and $\xi$ are the Higgs and Goldstone bosons.

Inserting (\ref{higgs}) in (\ref{kin44}) we obtain
\eqb 
{\cal L} _{eff}&=& \frac{1}{2} (\partial h)^2 - \frac{m^2}{2} (v+h)^2 - \frac{\lambda}{8}(v+h)^4 + \frac{{\tilde e}^2}{2} (v+h)^2 A_\mu A^{\mu} \nonumber 
\\ 
&& -\frac{1}{4}F_{\mu \nu} (A)F^{\mu \nu} (A) 
+
 \frac{(v+h)^2}{v} \partial^\mu \xi\left[\frac{1}{2 v}\partial_\mu \xi - \tilde{e} A_\mu\right].
\label{last1}
\eqf

Choosing the U-gauge, the last term in (\ref{last1}) vanishes, therefore 
\eqb 
{\cal L}_{eff} &=& \frac{1}{2} (\partial h)^2 - \frac{m^2}{2} (v+h)^2 - \frac{\lambda}{8}(v+h)^4 + \frac{{\tilde e}^2}{2} (v+h)^2 A_\mu A^{\mu}  -\frac{1}{4}F_{\mu \nu} (A)F^{\mu \nu} (A), \nonumber
\\ 
&=& \frac{1}{2} (\partial h)^2  - m^2 h^2 -\frac{1}{4}F_{\mu \nu} (A) F^{\mu \nu} (A)+  \frac{ m^2_A}{2} A_\mu A^{\mu} +  \frac{{\tilde e}^2}{2} h^2 A_\mu A^{\mu} +\cdots 
\label{last}
\eqf 
from where we see that
\bb 
m_A = {\tilde e} v,  \label{masses}
\ee 
is the dark photon mass\footnote{Instead of the Higgs mechanism, we can give mass to the dark photon by means of the Stueckelberg mechanism. Of course, the result would be the same.}. 

Already at this level we see that the effect of kinetic mixing not only amplifies the coupling constant but also the gauge boson mass. The same will happen when considering a more realistic model, where the dark $Z$ boson will have a rescaled mass.

Since the model under consideration has a redefined coupling constant, the cross section of any process involving this coupling must have an enhancement. To make this fact explicit, let us calculate the total cross section of a simple process. For example, consider FIG. 1, 
\begin{figure}[h]
\begin{center}
  \includegraphics[width=.4\textwidth]{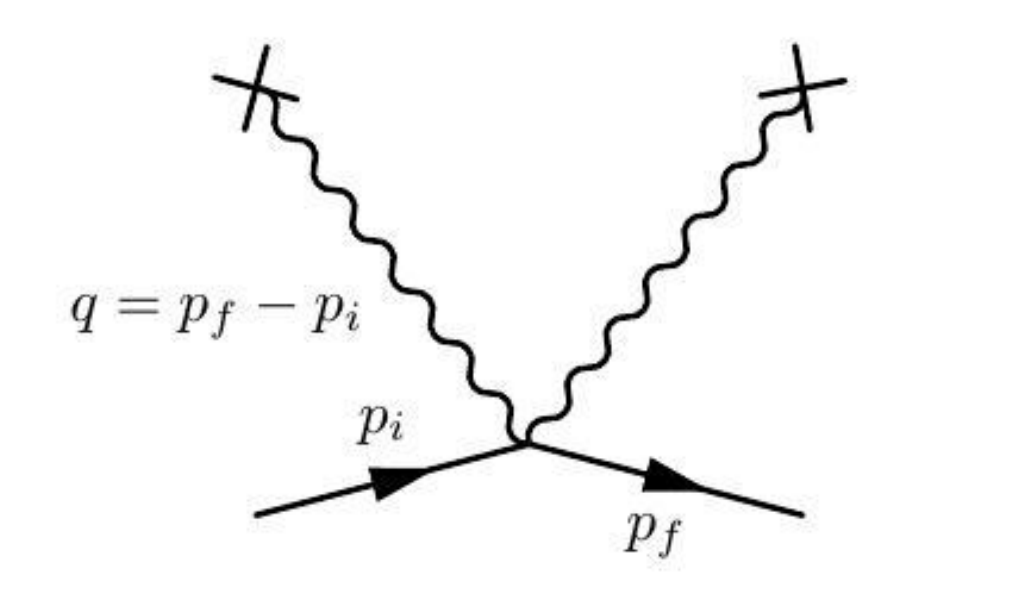}\\
  \caption{Higgs interacting with an external massive vector field.}\label{fig}
   \end{center}
\end{figure}

which is described by the term $$\frac{{\tilde e}^2}{2} h^2 A_\mu A^{\mu}, $$
where $A_\mu A^{\mu}$ is considered --as a simple application-- an external field. 

The $S$-matrix elements  are 
\begin{equation}
<f|S|i>=(-\frac{i}{2} {\tilde e}^2 \eta_{\mu \nu}) \frac{1}{\sqrt{2E_{f}2E_{i}}V} \int d^4 x e^{i(p_{f}-p_{i})\cdot x} A^{\mu}(x) A^{\nu}(x),
\end{equation}
with $V$ is a normalization volume and the product $A^{\mu}(x) A^{\nu}(x)$ is  a convolution integral in the Fourier space, {\it i.e.}
\begin{equation}
<f|S|i>=(-\frac{i}{2}{\tilde e}^2 \eta_{\mu \nu}) \frac{1}{\sqrt{2E_{f}2E_{i}}V} \int d^4 k A^{\mu}(q-k) A^{\nu}(k),
\end{equation}
 where $q=p_{f}-p_{i}$ is the transferred momentum. 

As the external field is massive, it can be expressed as

\begin{equation}
A^{\mu}(x)=\frac{ \alpha \,e^{-m_A r}}{r}  \longrightarrow       A^{\mu}(q)= \delta(q_{0}) \frac{2 \alpha }{{\bf q}^2+m^2_A} \eta^{\mu 0},
\end{equation}
where $\alpha$ is the source charge generating the external field. For simplicity, $\alpha$ will be normalized to 1.

Using these results, $<f|S|i>$ is 
\eqb 
<f|S|i>=-(i {\tilde e}^2) \frac{ \delta(q_{0})  }{\sqrt{E_{f}E_{i}}V} \int d^3 k  \frac{1}{(|{\bf q}-{\bf k}|^2+m^2_A)}    \frac{1}{({\bf k}^2+m^2_A)} \label{amp12}.
\eqf

In order to compute (\ref{amp12}) we use a Feynman parameterization and we get  
\begin{equation}
<f|S|i>=-(i {\tilde e}^2) \frac{ 2\pi^2 \delta(q_{0})  }{\sqrt{E_{f}E_{i}}V} \frac{1}{|{\bf q}|} \arcsin \frac{|{\bf q}|}{\sqrt{{\bf q}^2+4m_A^2}}.
\end{equation}

The differential cross section is given by
\begin{equation}
d\sigma=\frac{1}{|{\bf j_{in}}|} \frac{|S_{fi}|^2}{T} \frac{Vd^3 p_{f}}{(2\pi)^3},
\end{equation}
where  ${\bf j_{in}}={\bf v_{i}}/V$ is the incoming current and $T$ is a time normalization parameter \cite{drell}.

Collecting everything we get 
\begin{equation}
\frac{d\sigma}{ d\Omega_{f}}=\frac{{\tilde e}^4 }{4}   \left| i  \frac{1}{2|{\bf p}| \sin\frac{\theta}{2}}  \arcsin \frac{ \sin\frac{\theta}{2}}{\sqrt{\sin^2\frac{\theta}{2}+\frac{m^2_A}{|{\bf p}|^2}} }\right|^2= |f(\theta,\phi)|^2, \label{sec0}
\end{equation}
where the term  between $|...|^2$ is just the scattering amplitude and $|{\bf p}| =|{\bf p}_f|= |{\bf p}_i|$. 

The calculation of the total annihilation cross section is most easily done by using the optical theorem, namely 
\begin{equation}
\sigma_T =\int  |f(\theta,\phi)|^2 d\Omega=\frac{4\pi}{|{\bf p}|}{\cal I}\mbox{m} \left[f(\theta=0)\right], 
\end{equation}
and with the help of the identity
\begin{equation}
 \lim_{x \to 0} \frac{1}{x} \arcsin \frac{x}{\sqrt{x^2+a^2}}= \frac{1}{a}, 
\end{equation}
the total cross section becomes 
\begin{equation}
\sigma_T=\frac{\pi \tilde{e}^4 }{4|{\bf p}|\,m_A }, 
\end{equation}
which shows the Sommerfeld enhancement at low energies for this particular process. 

As mentioned above, although this calculation was made for a particular interaction, FIG. 1, it is clear that
any other electromagnetic process in the dark sector will have an enhanced cross section since it will depend on the rescaled electric
charge. Accordingly, the conclusion that the  coupling to the visible photon induces an enhancement in the
interactions is also true for more complicated processes. 

To higher orders --i.g., $n$-th order--  the total cross section should be proportional to $(1-(\gamma e_B e_A)^2)^{-n/2}$ and therefore  negligibly small in comparison to the tree level processes. 

\section{Extension to $SU(3)\times SU(2)\times U(1)$}

In order to extend these results to the case $SU(3) \times SU(2)\times U_Y(1)$, one proceed as follows; first, we consider the dark non abelian fields $G_\mu =G^a_\mu \lambda^a$ and $A_\mu= A^a_\mu \tau^\alpha$, where $\lambda^a$ and $\tau^\alpha$ are the generators of $SU(3)$
 and $SU(2)$ respectively, while the dark abelian field will be denoted by $B_\mu$.  

Following the above ideas, the non abelian fields will remain untouched, but the abelian field dynamics will be modified by extending the gauge group to $SU(3)\times SU(2)\times U_Y(1)\times U_{Y'} (1)$, where $U_{Y'} (1)$ is associated to the visible photon $B'_\mu$. Accordingly, consider the Lagrangian 
\bb
{\cal L} = {\cal L}_{\text{matter}} + {\cal L}_{\text{gauge}}, \label{12}
\ee 
where $ {\cal L}_{\text{matter}} $ denotes all matter fields in the dark sector, including a dark Higgs field and their interactions with the gauge fields. 

We are now interested in the gauge sector. Similarly to the above model, we include the visible field $B'_\mu$, coupled to $B_\mu$ through a kinetic mixing. The Lagrangian gets diagonalized 
\eqb 
{\cal L}_{\text{gauge}} &=&   -\frac{1}{4} \text{tr} F^2 (G) - \frac{1}{4} \text{tr} F^2 (A) -\frac{1}{4{{g'}^2}}F^2 (B) - \frac{1}{4{{g''}^2}}F^2 (B')+\frac{\gamma}{2}F (B) F(B') \nonumber
\\ 
&=&-\frac{1}{4} \text{tr} F^2 (G) - \frac{1}{4} \text{tr} F^2 (A) -\frac{1}{4{\tilde g}}F^2 (B) -\frac{1}{4g''^2}F^2({\tilde B}), 
\label{gau}
\eqf
where the coupling constant $g'$ gets redefined
\bb
{\tilde g}=\frac{g'}{\sqrt{1-(g' {g''} \gamma)^2}} \equiv \frac{g'}{\sqrt{1-\chi^2}}, \label{res}
\ee 
and the $\tilde{B}_\mu$ field is defined as
\[
{\tilde B}_\mu = B'_\mu -\gamma g''^2 B_\mu.
\]

From the path integral formulation, the field ${\tilde B}$ is decoupled from the dynamics of the other fields and the only relic of the kinetic mixing emerges from the rescaled coupling (\ref{res}). By redefining $B_\mu$ in (\ref{gau}) as
\[
\frac{B_\mu}{{\tilde g}} \to  {\cal B}_\mu.
\]
the covariant derivative turns out to be 
\bb
D_\mu =  \partial_\mu  -i g_S G_\mu  - i gA_\mu - i {\tilde g} {\cal B}_\mu  \label{covariant}
\ee 
and therefore  the dark sector becomes $SU(3)\times SU(2) \times U(1)$ with an effective hypercharge. 

We can now follow the lines of the Higgs mechanism in the usual way. The mass of the dark Z boson gets rescaled as 
\bb
M_Z \to  \frac{M_Z}{\sqrt{1-\chi^2}} \equiv M_{Z'}, 
\ee
as opposite to the masses of the $W$ which remains unchanged. Note that $M_{Z'}>M_Z$.

As a final remark, due to the redefinition of the $Z$ mass, the Fermi coupling constant $G_F$ also gets rescaled. Since $g$, the coupling constant of the weak sector, does not change, $G_F$ gets redefined
$$ G_F = \frac{g^2}{M^2_Z}\rightarrow G'_F = \frac{g^2}{M^2_{Z'}},$$ 
which implies  
\bb
\frac{G'_F}{G_F} = 1 -\chi^2 <1. \label{cota2}
\ee
This agrees with some phenomenological considerations \cite{santa}.

\section*{Acknowledgments}
 
This work was supported by grants from CONICYT-21140036 (D.C.), FONDECYT 1130020 (J.G.) and  11130083 and 7912010 (M.P.).

\end{document}